# Tunable negative thermal expansion related with the gradual evolution of antiferromagnetic ordering in antiperovskite manganese nitrides $Ag_{1-x}NMn_{3+x}$ (0≤x≤0.6)

**J. C. Lin**,[1] **P. Tong**,[1,a)] **W. Tong**,[2] **S. Lin**,[1] **B. S. Wang**,[1] **W. H. Song**,[1] **Y. M. Zou**,[2] and **Y. P. Sun**[2,1,3,b)]

[1] *Key Laboratory of Materials Physics, Institute of Solid State Physics, Chinese Academy of Sciences, Hefei 230031, People's Republic of China*

[2] *High Magnetic Field Laboratory, Chinese Academy of Sciences, Hefei 230031, People's Republic of China*

[3] *Collaborative Innovation Center of Advanced Microstructures, Nanjing University, Nanjing 210093, People's Republic of China*

a) E-mail: tongpeng@issp.ac.cn

b) E-mail: ypsun@issp.ac.cn

The thermal expansion and magnetic properties of antiperovskite manganese nitrides $Ag_{1-x}NMn_{3+x}$ were reported. The substitution of Mn for Ag effectively broadens the temperature range of negative thermal expansion and drives it to cryogenic temperatures. As $x$ increases, the paramagnetic (PM) to antiferromagnetic (AFM) phase transition temperature ($T_N$) decreases. At $x$~0.2, the PM-AFM transition overlaps with the AFM to glass-like state transition. Above $x$=0.2, two new distinct magnetic transitions were observed: one occurs above room temperature from PM to ferromagnetic (FM), and the other one evolves at a lower temperature ($T^*$) below which both AFM and FM orderings are involved. Further electron spin resonance measurement suggests that the broadened volume change near $T^*$ is closely related with the evolution of $\Gamma^{5g}$ AFM ordering.

## I. INTRODUCTION

Materials with negative thermal expansion (NTE) which can compensate the normal thermal expansion have potential applications in many fields where precise control of the thermal expansion is indispensible.[1-7] However, most NTE materials reported so far are ceramic, which restricts their applications in relevant fields where metallic features (e.g., high electrical or thermal conductivity, high stiffness, etc.) are desirable.[7] The recent reports of NTE in antiperovskite manganese nitrides $ANMn_3$ (A represents transition metals or semiconducting elements) have attracted considerable attention because of the isotropic cubic lattice, tunable thermal expansion coefficient, wide temperature span, and the metallic properties.[8-15] The NTE in $ANMn_3$ compounds mainly originates from the broadening of the sharp volume contraction (i.e., the magneto-volume effect, MVE) upon heating at the paramagnetic (PM) to antiferromagnetic (AFM) phase transition by partial doping on A site.[8,11,13,15] The element doping on A site usually triggers an increase in the PM-AFM transition temperature ($T_N$) of $ANMn_3$.[8,9,11,16,17] Consequently, the NTE occurring at low temperatures is relatively rare in doped $ANMn_3$.[10] In practice, however, the NTE at cryogenic temperatures is as equivalently important as that at room temperature.[18,19] In previous studies, the exploration of NTE has been focused on a few antiperovskite compounds that display large MVE, such as slightly doped $CuNMn_3$ with Ge or Sn on Cu site, $ZnNMn_3$ and $GaNMn_3$.[8-14,16,20-22] However, very few studies have examined the NTE based on $AgNMn_3$ which also exhibits considerably large MVE around $T_N$ (~276 K).[23] Here, we report the observation of NTE at cryogenic temperatures in $Ag_{1-x}NMn_{3+x}$ by substituting Ag with Mn. The volume change at $T_N$ is quite sharp below $x \leq 0.3$, but it broadens when $x > 0.3$. Moreover, our electron spin resonance (ESR) result suggests the broadening of volume change (i.e., NTE) is closely related to the slow development of AFM ordering. Additionally, the evolution of magnetic phase transitions with temperature and $x$ is mapped.

## II. EXPERIMENTS

Polycrystalline samples of $Ag_{1-x}NMn_{3+x}$ ($0 \leq x \leq 0.6$) were prepared using solid-state reaction. Powders of Ag (4N), Mn (4N) and homemade $Mn_2N$ were mixed according to appropriate ratios, pressed into pellets, sealed in evacuated quartz tubes ($\sim 10^{-6}$ torr) and sintered at 750 ºC for 3 days, followed by annealing at 800 ºC for 5 days. After quenching the tubes to room temperature, the products were ground carefully, pressed into pellets, sealed in evacuated tubes and sintered at 800 ºC for extra 8 days. X-ray diffraction showed the obtained samples exhibit the antiperovskite structure (group symmetry: Pm3m) without detectable second phase. The magnetic measurements were performed either on a Quantum Design Superconducting Quantum Interference Device (QD-SQUID) magnetometer ($1.8K \leq T \leq 400K$, $0 \leq H \leq 50kOe$) or on a vibrating sample magnetometer with an oven attached to a Quantum Design Physical Property Measurement System (QD-PPMS). Linear thermal expansion $\Delta L/L(365K)$, where $\Delta L$ is the linear change of the lattice compared with the value at 365K, i.e., $\Delta L = L(T) - L(365K)$, was measured using a strain gauge (KYOWA, type KFL). High-pure copper (purity 99.99%) was used as a reference whose thermal expansion coefficient was known.[24] The electron spin resonance (ESR) spectra were recorded using an X-band Bruker EMX plus 10/12 cw spectrometer operating at 9.4 GHz.

## III. RESULTS AND DISCUSSION

Figure 1 shows the linear thermal expansion $\Delta L/L(365K)$ for $Ag_{1-x}NMn_{3+x}$ ($0 \leq x \leq 0.6$). For $x=0$, the lattice contracts below $T_N$ (279K) with $\Delta T=8K$, which is in line with data recently reported by Takenaka et al.[23] The thermal expansion coefficient $\alpha = \Delta L/[L(365K)\Delta T]$ is estimated to be $-44.5 \times 10^{-6}$/K. For $x=0.3$, the temperature span for the lattice contraction is increased to 20K (213K-233K) and the magnitude of $\alpha$ ($-57.0 \times 10^{-6}$/K) is slightly larger than that of the parent compound. As $x$ increases further, the volume change keeps broadening, leading to the reduced magnitudes of $\alpha$. The $\alpha$ value is $-16.7 \times 10^{-6}$/K (175K-212K), $-5.1 \times 10^{-}$

$^{-6}$/K (135K-185K) and -0.48×10$^{-6}$/K (5K-87K) for $x$=0.4, 0.5 and 0.6, respectively. In addition, the normal positive thermal expansion of lattice contraction below the temperature region is also suppressed when $x$ increases. As a result, the $\alpha$ value is only 1.65×10$^{-6}$/K below 130K for $x$=0.5. The small $\alpha$ values below the liquid nitrogen temperature (~77K) observed when $x$=0.5 and 0.6 may have potential applications in cryogenic engineering.

Figure 2 presents the temperature dependent magnetization $M$(T) of Ag$_{1-x}$NMn$_{3+x}$(0≤$x$≤0.6) measured at $H$=100 Oe under both zero-field-cooled (ZFC) and field-cooled (FC) modes. The $M$(T) curves for $x$≤0.2 are shown in Fig. 2(a). For $x$=0, the ZFC $M$(T) curve exhibits a peak around 100K on cooling. The large divergence between the ZFC and FC curves at low temperatures resembles a glass-like behavior. As shown in Fig. S1,[25] the dynamic magnetic susceptibility suggests a spin-glass (SG) state below the spin freezing temperature $T_g$ (~100K). The presence of SG state is due to the competing FM and AFM interactions.[26] Such a SG state is maintained up to $x$=0.2 whose $M$(T) shape is the same as that of $x$=0. The PM-AFM transition at $T_N$ is evidenced by a kink around 283K on the ZFC $M$(T) curve as shown in the inset of Fig. 2(a). This accords with the reported values of $T_N$ varying from 276K to 300K[23,26] and the temperature where the sharp MVE occurs (Fig. 1). As presented in Fig. 2(a), $T_N$ shifts towards lower temperature when $x$ increases, while $T_g$ shifts towards higher temperature. $T_N$ and $T_g$ overlap at $x$~0.2. The kink-like feature in ZFC $M$(T) curve is no longer visible for samples with $x$≥0.2. Instead, a PM-FM like magnetic transition can be observed at high temperature for $x$>0.2, as displayed in Fig. 2(b). The PM-FM transition temperature ($T_C$) increases as $x$ rises. The ZFC $M$(T) curve drops at certain temperatures (noted as $T^*$) upon cooling, indicating another magnetic transition. $T^*$ decreases as $x$ increases. In addition, the divergence between the ZFC and FC $M$(T) curves is evident in samples when $x$>0.3. However, as shown in Fig. S2,[25] ac magnetic susceptibility measurement suggests the spins are ordered below $T^*$ rather than randomly frozen in a SG state. More interestingly, the magnetic transition at $T^*$ becomes broadened when $x$ increases. This broadening, which basically accords with the NTE temperature span, is more clearly demonstrated by the *dM/dT* curves as shown in the inset of Fig. 2(b). It implies that the lattice and magnetism are closely correlated near $T^*$ when $x>0.3$.

Figure 3(a) presents the magnetization isotherms *M*(H) for $Ag_{1-x}NMn_{3+x}$ ($0\leq x\leq 0.6$) at 250K. For $x<0.2$ (e.g., $x=0$ and 0.05), the *M*(H) value is small and shows a linear temperature dependence, which is consistent with $\Gamma^{5g}$ AFM state.[26] However, when $x>0.2$ (e.g., $x=0.3-0.6$), the magnetization at 250K increases immediately in external magnetic field, giving rise to a very weak coercivity ($H_C$). Moreover, the *M*(H) is almost saturated at 45 kOe, particularly for samples with high *x* values, e.g., $x=0.6$. This is in line with the FM-like ZFC *M*(T) curves in Fig. 2(b). The *M*(H) measured at 5K is presented in Fig. 3(b), which differs essentially from those measured at 250K. At first, the *M*(H) is not saturated at 45 kOe. Secondly, large hysteresis loops are clearly observable in the *M*(H) curves. As displayed in the right inset of Fig. 3(b), $H_C$ is very weak below $x=0.2$ (~20 Oe), but remarkably increases at higher doping levels. For $x=0.6$, $H_C$ is about 4 kOe. Meanwhile, as seen in the left inset of Fig. 3(b), the magnetization at 45 kOe increases initially when *x* rises up to 0.3 and then reduces when *x* increases further. The reduced magnetization at 45 kOe for $x>0.3$ may be indicative of the existence of AFM ordering which agrees with the drop in ZFC *M*(T) curves at $T^*$ upon cooling. The AFM coupling would drag the FM spins irreversibly, and thus cause the remarkably enhanced coercivity for $x>0.3$.[27]

Based on the magnetic transition temperatures ($T_N$, $T_g$, $T_C$ and $T^*$) discussed above, a phase diagram for $Ag_{1-x}NMn_{3+x}$ ($0\leq x\leq 0.6$) was plotted in Fig. 4. Meanwhile, the temperature zone where the lattice contraction appears is highlighted in the phase diagram. The $x=0.2$ sample turns out to be a critical composition since the four transition temperatures merge here. For $x\leq 0.2$, the NTE zone is well localized just below the $T_N(x)$ line, showing MVE rather than NTE. At higher doping levels, the NTE zone becomes wider when *x* increases and its onset temperature coincides with the $T^*(x)$ line. This finding unambiguously demonstrates that the broadening of volume change correlates with the evolution of the magnetic ordering below $T^*$.

To further shed light on the relation between the NTE and the magnetic transition around $T^*$, temperature dependent ESR was measured from 60K to 296K for $x$=0.5. The derivative ESR spectra $dP/dH$ is shown in Fig. 5. Here, the $dP/dH$ data were shifted accordingly and data at low temperatures were multiplied by an appropriate factor so that the change of the line shape with temperature could be easily observable. At 296K, which is much lower than $T_C$ (~ 440K), the spectrum shows a single resonance. The resonant field (~ 2.3 kOe) is significantly smaller than the usual PM resonant field of 3.35kOe, indicating a FM state. Upon cooling, the $dP/dH$ spectrum shifts towards lower magnetic fields due to the growth of the internal field.[28] Meanwhile, the spectrum becomes distorted below 190K which is close to the magnetic transition at $T^*$ (182K). Synchronously, a new resonance distinguishable from the original one appears at a higher resonant field. This high-field resonance is attributable to AFM coupling which requires a stronger magnetic field to generate ESR resonance than the PM one.[29,30] However, such an AFM coupling should be short-range (SR) ordered, because the strong AFM spin coupling in the long-range ordered case demands a resonance field which is much stronger than the magnitude of ~10 kOe used in a X-band ESR.[31] As the temperature further decreases, the overall spectrum intensity keeps declining, while the AFM resonant peak shifts quickly toward higher magnetic fields and finally disappears below $T$~120K. On the other hand, the FM resonance signal, which is demonstrated by the low-field resonant peak (marked by the asterisks in Fig. 5), still presents below the onset temperature of the SR-AFM ordering. The ESR result confirms the conclusion deduced from Fig. 3(b). Namely, the FM ordering observed in ESR explains the overall FM-like behavior of the isothermal $M$(H)s, whereas the AFM ordering revealed by the ESR data confirms the reduced magnetization and enhanced $H_C$ at high doping levels.

The doubly integrated intensity (DIN) is plotted in the inset of Fig. 5. It shows a remarkable reduction of DIN below ~200K due to the onset of AFM ordering.[32] The reduction of DIN with the decrease of temperature is gradual, which is in good accordance with the well-dispersed transition region around $T^*$ as presented in the inset of Fig. 2(b). More importantly, as illustrated in the inset of Fig. 5, the NTE temperature span accords precisely with the temperature range where the AFM ordering grows gradually. Due to the geometrical frustration in terms of AMF coupling, both $\Gamma^{5g}$ and $\Gamma^{4g}$ types of triangular AFM configurations were observed in the antiperovskite manganese nitrides. However, only the $\Gamma^{5g}$ type is related to the lattice expansion compared with the PM state, and thus is believed to be a prerequisite for the broadening in the magnetic/volume change, namely the NTE.[23] So, the AFM signal observed here can be reasonably attributable to the $\Gamma^{5g}$ type. Although it is commonly acknowledged that the gradually developing $\Gamma^{5g}$ AFM moment is relevant to the gradual volume expansion,[13,15,33] how the lattice change correlates with the magnetic ordering is not well-understood yet. Our ESR findings may add new information to understand the broadened MVE. When it is below $T^*$, the locally ordered $\Gamma^{5g}$ AFM interactions may lead to a local lattice expansion. Upon cooling, the range of the $\Gamma^{5g}$ AFM interactions becomes extended. Accordingly, the regions with expanded lattice would become larger and coalesce with each other, leading to a continuous expansion of the average lattice. Eventually, the average lattice constant reaches at a maximum value when the $\Gamma^{5g}$ AFM interactions become long-range ordered.

The locally ordered AFM coupling may originate from the local structure distortion as reported previously,[15,34,35] which can locally relieve the geometrical frustration. As Mn doping level ($x$) increases, the enhanced local structure distortion would relax the structural/magnetic transition[15,34,35] and thus lead to the broadening of MVE above $x$=0.3. Chemical doping probably introduces local distortion because of the spatially compositional fluctuations in such polycrystalline solid solutions. However, not all the dopants, but only those which disturb the $\Gamma^{5g}$ AFM state are effective in broadening the MVE.[23] To further understand the mechanism of MVE broadening due to Mn doping, the spin configurations for FM and AFM ordering as well as the local structure need to be clarified for $Ag_{1-x}NMn_{3+x}$. Even so, the current findings suggest that ESR can provide a potential new tool for probing the coupling between the magnetic states and the gradual volume change, which is significantly crucial for understanding the NTE phenomenon in the antiperovskite manganese nitrides.

## IV. CONCLUSION

In summary, the thermal expansion and magnetic properties were studied in $Ag_{1-x}NMn_{3+x}$ ($0 \leq x \leq 0.6$). For $x<0.2$, when $x$ increases, $T_N$ decreases, whereas $T_g$ shifts upwards. For $x>0.2$, upon cooling, the samples undergo two magnetic transitions successively: the PM-FM transition at $T_C$ and the transition at $T^*$ from FM state to a complex state in which AFM and FM orderings coexist. As $x$ increases, $T_C$ increases, while $T^*$ decreases. The lattice contraction, which happens at $\sim T_N$ for $x<0.2$ and $\sim T^*$ for $x>0.2$, becomes broadened and shifts towards cryogenic temperatures when $x>0.3$. Furthermore, revealed by ESR measurement for $x=0.5$, the NTE temperature span covers well the temperature range where the $\Gamma^{5g}$ type AFM ordering evolves from short-range to long-range as the temperature decreases.

## ACKNOWLEDGMENTS

This work was supported by the National Key Basic Research under Contract Nos. 2011CBA00111 and the National Natural Science Foundation of China under Contract Nos. 51322105, 11174295, 51301167, 51171177 and 91222109, and the Foundation of Hefei Center for physical science and technology under Contract No. 2012FXCX007. The first author thanks Dr. C. Sun from University of Wisconsin-Madison for her assistance in editing the revised manuscript.

**Figures:**

FIG.1. Linear thermal expansion $\Delta L/L$(365K) for $Ag_{1-x}NMn_{3+x}$ ($0\leq x\leq 0.6$). Solid lines are linear fitting to the data and the related coefficients of linear thermal expansion are shown.

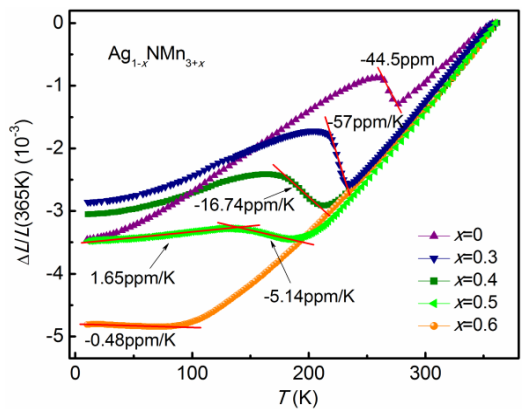

Lin et al., Figure 1

FIG. 2. The magnetization $M$(T)s measured at both zero-field-cooling (ZFC) (solid symbols) and field-cooling (FC) (open symbols) modes for $Ag_{1-x}NMn_{3+x}$ with $x \leq 0.2$ (a) and $x \geq 0.3$ (b). Inset in (a) shows an enlargement of the high-temperature ZFC $M$(T) for $x \leq 0.2$. Inset in (b) shows the $dM/dT$ vs $T$ curve around $T^*$ for $x \geq 0.3$. The shifts of the magnetic transitions ($T_N$, $T_C$, $T_g$ and $T^*$) with $x$ are indicated by the arrows.

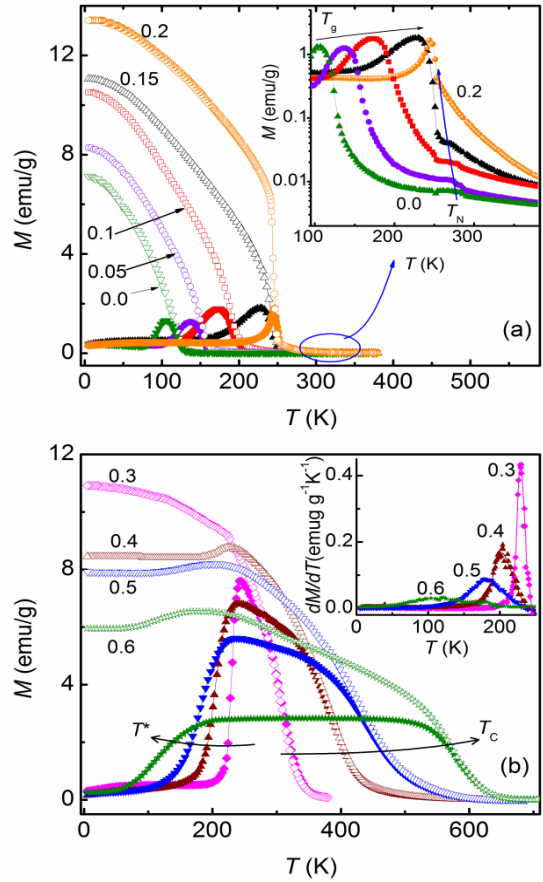

Lin et al., Figure 2

FIG. 3. The isothermal magnetization $M$(H) Ag$_{1-x}$NMn$_{3+x}$($0 \leq x \leq 0.6$) at 250K (a) and at 5K (b). Left and right insets in (b) show the $x$ dependent magnetization at 45 kOe ($M_{45kOe}$) and coercive field ($H_C$), respectively.

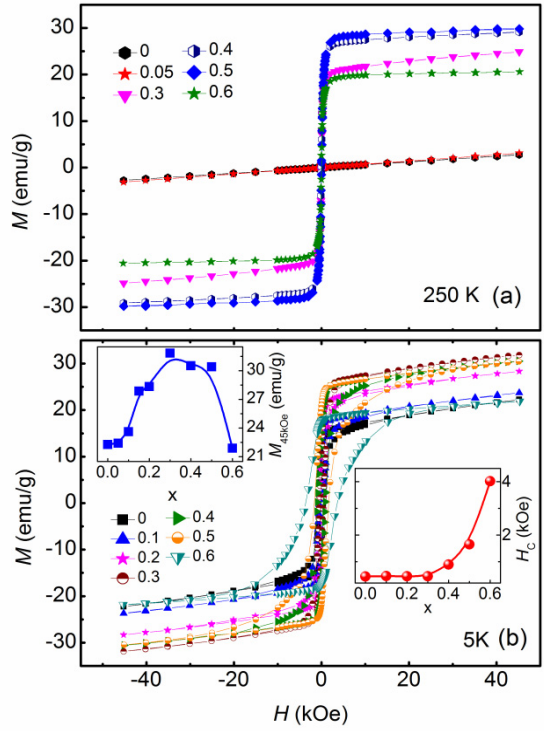

Lin et al., Figure 3

FIG. 4. Phase diagram for $Ag_{1-x}NMn_{3+x}$ ($0 \leq x \leq 0.6$). The values of the magnetic transition temperatures $T_N$, $T_C$, $T_g$, $T^*$ and the magnetic states are determined by the magnetic susceptibility measurements. The temperature zone where the lattice contracts on heating is highlighted. The onset and the end temperatures ($T_{NTE}$) for the lattice contractions are shown as well.

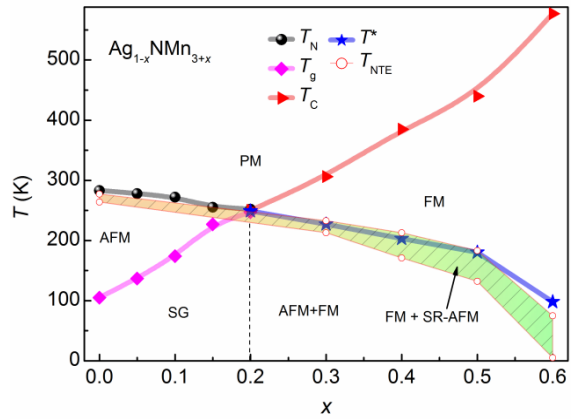

Lin et al., Figure 4

FIG. 5. ESR spectra, $dP/dH$, as a function of magnetic field at different temperature for $Ag_{1-x}NMn_{3+x}$ with $x=0.5$. The evolution of the two resonances are indicated by "*" and "^". The ESR spectra under different temperature were shifted accordingly and the data at low temperatures were multiplied by an appropriate factor so that the evolution of the line shape with temperature could be easily observable. Inset shows the doubly integrated intensity (DIN) of the ESR spectra plotted as log(DIN) vs $T$. The $T^*$ deduced from the magnetic susceptibility is marked and the temperature zone for negative thermal expansion is highlighted.

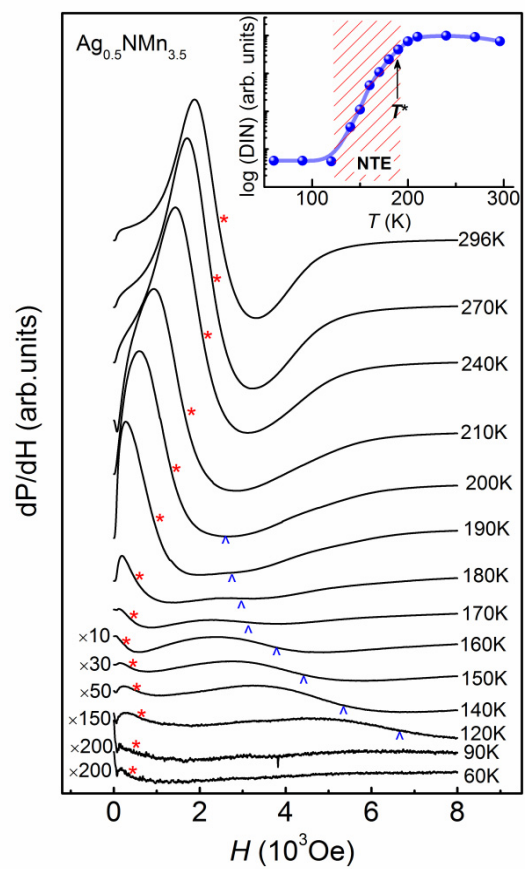

Lin et al., Figure 5